\documentclass[conference]{IEEEtran}
\usepackage{cite}
\usepackage{amsmath,amssymb,amsfonts}
\usepackage{algorithmic}
\usepackage{graphicx}
\usepackage{textcomp}
\usepackage{xcolor}
\usepackage{url}
\usepackage{hyperref}

\def\BibTeX{{\rm B\kern-.05em{\sc i\kern-.025em b}\kern-.08em
    T\kern-.1667em\lower.7ex\hbox{E}\kern-.125emX}}

\begin{document}

\title{Stealth and Evasion in Rogue AP Attacks: An Analysis of Modern Detection and Bypass Techniques}

\author{\IEEEauthorblockN{Kaleb Bacztub, Braden Vester, Matteo Hodge, Liulseged Abate}
\IEEEauthorblockA{\textit{Department of Computer Information Technology} \\
\textit{Purdue University}\\
Indianapolis, IN, USA \\
kbacztu@purdue.edu, bjvester@purdue.edu, mlhodge@purdue.edu, lmabate@purdue.edu}
}

\maketitle

\begin{abstract}
Wireless networks act as the backbone of modern digital connectivity, making them a primary target for cyber adversaries. Rogue Access Point (AP) attacks, specifically the ``Evil Twin'' variant, enable attackers to clone legitimate wireless network identifiers to deceive users into connecting. Once a connection is established, the adversary can intercept traffic and harvest sensitive credentials. While modern defensive architectures often employ Network Intrusion Detection Systems (NIDS) to identify malicious activity, the effectiveness of these systems against Layer 2 wireless threats remains a subject of critical inquiry. This project aimed to design a stealth-capable Rogue AP and evaluate its detectability against Suricata, an open-source NIDS/IPS. 

The methodology initially focused on a hardware-based deployment using Raspberry Pi platforms but transitioned to a virtualized environment due to severe system compatibility issues. Using \textit{Wifipumpkin3}, the research team successfully deployed a captive portal that harvested user credentials from connected devices. However, the Suricata NIDS failed to flag the attack, highlighting a significant blind spot in traditional intrusion detection regarding wireless management frame attacks. This paper details the construction of the attack, the evasion techniques employed, and the limitations of current NIDS solutions in detecting localized wireless threats.
\end{abstract}


\section{Introduction}
\label{sec:introduction}
The ubiquity of Wi-Fi networks has revolutionized connectivity but simultaneously introduced a vast and often insecure attack surface. As organizations and individuals rely increasingly on wireless infrastructure, the ``Evil Twin'' attack persists as a potent threat vector. In this scenario, an adversary broadcasts a malicious network SSID (Service Set Identifier) that mimics a legitimate access point, exploiting the automatic connection behaviors of modern devices.

The primary objective of this project was to answer a critical defensive question: Can a sophisticated Rogue AP, employing modern evasion techniques, bypass standard detection mechanisms? The initial proposal envisioned a hardware-centric approach utilizing Raspberry Pi devices to physically mimic network infrastructure. However, the project evolved significantly during the implementation phase. Technical hurdles regarding driver support for the Raspberry Pi 5 necessitated a shift toward a fully virtualized architecture.

This paper explores the theoretical underpinnings of Man-in-the-Middle (MitM) attacks facilitated by Rogue APs and analyzes the efficacy of evasion techniques such as MAC address randomization. Furthermore, it evaluates the performance of the Suricata Intrusion Detection System (IDS) when vetted against a localized wireless attack. The findings suggest that while NIDS are robust against payload-based attacks, they often lack the visibility required to detect Layer 2 wireless anomalies without specialized wireless sensors or Wireless Intrusion Prevention Systems (WIPS).

\section{Background}
\label{sec:background}

\subsection{The Ubiquity of Wi-Fi and Trust Exploitation}
Wi-Fi has become a utility akin to electricity in public spaces. This prevalence has fostered a culture of implicit trust among users, who frequently connect to familiar SSIDs without verifying the integrity of the broadcasting hardware. The classic ``Coffee Shop'' scenario illustrates this vulnerability: a user connects to an open network named ``Free Coffee Wi-Fi,'' unaware that the signal emanates from an attacker's laptop rather than the shop's router. This exploitation of trust is the foundation of the Rogue Access Point (RAP) attack.

\subsection{Rogue Access Points and Evil Twins}
A Rogue AP is any unauthorized access point connected to a network. An Evil Twin is a Rogue AP specifically designed to impersonate a legitimate AP. By cloning the SSID and often the MAC address (BSSID) of a valid network, the attacker forces the victim's device to connect to the stronger signal provided by the rogue device. Once connected, the attacker achieves a Man-in-the-Middle (MitM) position, enabling the interception of data, DNS spoofing, and the injection of malicious content, such as fake captive portals.

\subsection{Detection vs. Evasion}
Defenders typically rely on WIPS or NIDS to identify these threats. Detection methods often look for signal strength anomalies, sequence number gaps in management frames, or duplicate MAC addresses. To counter this, attackers employ evasion techniques. MAC address randomization prevents simple blacklisting, while probe response manipulation allows the Rogue AP to answer specific requests from victim devices, mimicking the behavior of enterprise-grade hardware.

\section{Methodology}
\label{sec:methodology}

The project development lifecycle was divided into two distinct phases. Phase I attempted a physical, field-deployable hardware implementation, while Phase II transitioned to a virtualized environment to resolve insurmountable hardware compatibility issues.

\subsection{Phase I: Hardware-Based Architecture}
The initial architectural design aimed to create a realistic, portable attack platform capable of being deployed in physical environments such as coffee shops or campus study areas. This setup utilized distinct physical hardware for both the legitimate and rogue infrastructure.

\subsubsection{Legitimate Infrastructure}
To establish a control environment, a ``Legitimate Access Point'' (AP) was constructed using a Raspberry Pi 3B+ running the \textit{RaspAP} software suite. This device acted as a fully functioning router, providing DHCP and DNS services, and was physically connected to an upstream internet source. It broadcasted the target SSID to establish a baseline of connectivity for the victim device.

\subsubsection{Rogue Infrastructure}
The attack platform, designated as the ``Rogue Access Point'' (RAP), was built on a Raspberry Pi 5 running the Kali Linux ARM architecture. To ensure high-power transmission and monitoring capabilities, an Alfa AWUS036ACH wireless adapter was attached to the Pi 5.

The attack flow was designed as follows:
\begin{enumerate}
    \item The RAP identifies the target network and forces deauthentication of connected clients.
    \item The Victim Machine, disconnected from the Legitimate AP, automatically connects to the RAP, which is broadcasting a stronger signal with an identical SSID.
    \item The RAP presents a captive portal to harvest credentials before allowing internet pass-through.
\end{enumerate}

\begin{figure}[htbp]
\centering
\includegraphics[width=\linewidth]{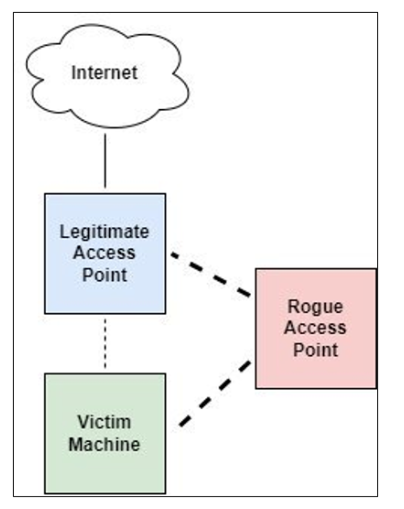}
\caption{Phase I Hardware Architecture. The Rogue AP (Red) was hosted on a Raspberry Pi 5, attempting to intercept traffic between the Victim (Green) and the Internet, bypassing the Legitimate AP (Blue).}
\label{fig:hardware_setup}
\end{figure}

\subsubsection{Failures and Limitations}
Despite the theoretical soundness of this architecture, the implementation faced critical failures due to the ARM architecture of the Raspberry Pi 5 and kernel-level driver incompatibilities.

\begin{table}[htbp]
\caption{Phase I: Tool Compatibility Failures }
\begin{center}
\begin{tabular}{|p{1.5cm}|p{1.5cm}|p{4cm}|}
\hline
\textbf{Technology} & \textbf{Outcome} & \textbf{Reason for Failure} \\
\hline
Nginx & Plausible & Encountered configuration issues when attempting to capture POST requests from the captive portal. \\
\hline
Pyngrok & Failed & Unable to properly install or tunnel traffic due to improper system architecture (ARM64) support on the Pi 5. \\
\hline
SET (Social Engineering Toolkit) & Plausible & Failed due to a lack of documentation regarding captive portal integration and general design conflicts with the Kali ARM kernel. \\
\hline
Drivers & Critical Failure & The Realtek drivers for the Alfa AWUS036ACH were unstable on the Pi 5 kernel, preventing simultaneous Monitor Mode and AP broadcasting. \\
\hline
\end{tabular}
\label{tab:failures}
\end{center}
\end{table}

As detailed in Table \ref{tab:failures}, the primary bottleneck was hardware-software integration. The \textit{Social Engineering Toolkit} (SET) proved unreliable for this specific captive portal application, and \textit{Pyngrok} could not function on the ARM architecture. Most critically, the wireless drivers required for the Alfa adapter were unavailable or unstable for the specific Linux kernel version running on the Raspberry Pi 5. This prevented the creation of a stable master mode interface required for \textit{hostapd}.

\subsection{Phase II: Virtualized Architecture Transition}
To mitigate the hardware limitations, the project pivoted to a fully virtualized environment (Phase II). This shift allowed for granular control over system dependencies and kernel versions, ensuring that the necessary wireless tools functioned as intended.

\begin{figure}[htbp]
\centering
\includegraphics[scale=.75]{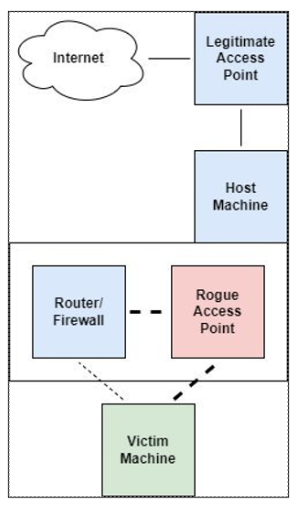}
\caption{Virtualized Experimental Architecture. The setup isolates the Legitimate AP (TP-Link TL-WR841N) from the attack infrastructure. The Rogue AP operates within a Kali Linux VM using USB passthrough for the Alfa adapter, while network traffic is monitored by Suricata on an Ubuntu host.}
\label{fig:virt_setup}
\end{figure}

The virtualized architecture (Fig. \ref{fig:virt_setup}) resolved the driver issues by utilizing a standard x64 Kali Linux Virtual Machine (VM) running on a host laptop. The specific components included:
\begin{enumerate}
    \item \textbf{Legitimate AP:} The Raspberry Pi was replaced with a dedicated TP-Link TL-WR841N router to strictly separate the ``safe'' network from the virtualized attack lab.
    \item \textbf{Attacker Machine:} A VMware virtual machine running Kali Linux. Crucially, the Alfa AWUS036ACH adapter was connected via USB passthrough, allowing the VM to directly control the physical radio while bypassing the host OS drivers.
    \item \textbf{IDS/NIDS:} An Ubuntu Server VM was introduced running Suricata. It was configured for inter-VLAN routing to monitor traffic flowing between the victim and the internet, attempting to detect the Rogue AP activity.
\end{enumerate}

\subsection{Attack Execution}
With the stable virtual environment, the attack was executed using \textit{Wifipumpkin3}, which replaced the failed tools from Phase I. The process followed these steps:
\begin{enumerate}
    \item \textbf{Target Identification:} Using \texttt{airodump-ng}, the team identified the target SSID and the channel of the legitimate AP.
    \item \textbf{Deauthentication:} A deauthentication attack was launched against the legitimate AP to force the victim client to disconnect.
    \item \textbf{Evil Twin Deployment:} \textit{Wifipumpkin3} was used to broadcast the clone SSID on the same channel.
    \item \textbf{Credential Harvesting:} Upon connecting to the Rogue AP, the victim was redirected via DNS spoofing to a fake login portal. Credentials entered were logged in cleartext by the attacker.
\end{enumerate}
\section{Results}
\label{sec:results}

\subsection{Attack Success}
\begin{figure}[htbp]
\centering
\includegraphics[width=\linewidth]{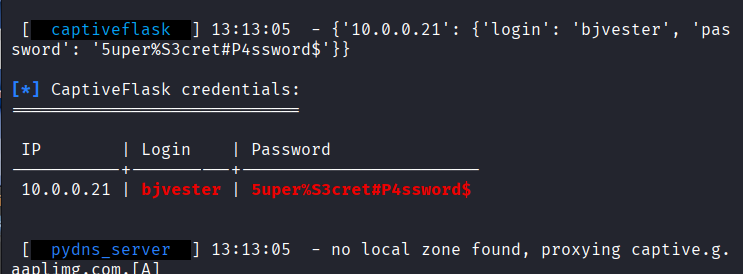}
\caption{Successful Credential Harvest via Wifipumpkin3. The terminal output shows the victim's IP address and the harvested login credentials.}
\label{fig:portal}
\end{figure}

The transition to the virtualized environment proved successful. The \textit{Wifipumpkin3} framework correctly handled the DHCP and DNS services required to route the victim to the captive portal.
\begin{itemize}
    \item \textbf{Connection:} The victim device successfully connected to the Rogue AP following the deauthentication flood.
    \item \textbf{Harvesting:} The cloned portal (Fig. \ref{fig:portal}) successfully captured the username and password, displaying them in the attacker's terminal.
\end{itemize}

\subsection{Detection Analysis}
A critical finding of this study was the failure of the configured NIDS to detect the attack. Suricata is a highly capable NIDS for analyzing IP traffic (Layer 3 and above). However, the Evil Twin attack fundamentally operates at Layer 2 (Data Link Layer) of the OSI model during the association phase.

The NIDS was monitoring the network flow, but because the Rogue AP created a separate network segment (a new subnet for the victim), the malicious traffic did not pass through the sensor in a way that triggered standard signatures. Unless a NIDS is specifically configured with a wireless interface in monitor mode and equipped with rules to analyze 802.11 management frames (such as Deauthentication packets or Beacon mismatches), it remains blind to the initial stages of a Rogue AP attack. This confirms limitations observed in similar studies regarding the need for specialized sensors for wireless monitoring.

\section{Future Work}
\label{sec:future}
The limitations identified in the detection phase open several avenues for future research:
\begin{itemize}
    \item \textbf{Advanced IDS/IPS Integration:} Future iterations of this project will involve building a dedicated Wireless Intrusion Prevention System (WIPS) capable of monitoring 802.11 management frames. Integrating tools like \textit{Kismet} with Suricata could bridge the gap between Layer 2 visibility and Layer 3 analysis.
    \item \textbf{Sophisticated Cloning:} While the captive portal used was effective, it was a template. Future work involves using \texttt{wget} or similar tools to clone dynamic, multi-factor authentication (MFA) pages to test user resilience against high-fidelity social engineering.
    \item \textbf{Live Environment Testing:} To better understand real-world efficacy, the team proposes conducting this attack within a live network under a strict Rules of Engagement (ROE) agreement, moving beyond the isolated lab environment.
    \item \textbf{Tool Comparison:} Testing alternative frameworks such as \textit{Wifiphisher} alongside \textit{Wifipumpkin3} to compare the effectiveness of different automated attack suites.
\end{itemize}

\section{Critical Analysis: The NIDS vs. WIDS Distinction}
\label{sec:discussion}

A retrospective analysis of the experimental results highlights a fundamental distinction in defensive architectures that was initially overlooked: the difference between a Network Intrusion Detection System (NIDS) and a Wireless Intrusion Detection System (WIDS).

The failure of Suricata to detect the Evil Twin attack was not indicative of a flaw in the detection software itself, but rather a misapplication of the tool within the experimental design. Suricata, operating as a NIDS, primarily inspects traffic at the Network Layer (Layer 3) and above. It analyzes IP packets, TCP handshakes, and application layer protocols to identify malicious signatures. However, the core mechanisms of a Rogue AP attack---specifically the Deauthentication floods and the broadcasting of spoofed Beacon frames---occur strictly at the Data Link Layer (Layer 2) within the 802.11 protocol standards.

Had the defensive environment incorporated a dedicated WIDS (such as Kismet or an enterprise-grade WIPS sensor), the attack would likely have been flagged immediately. A WIDS operates by monitoring the radio frequency (RF) spectrum directly, analyzing 802.11 management and control frames that a standard NIDS interface ignores. Specifically, a WIDS would have detected:
\begin{itemize}
    \item \textbf{Deauthentication Floods:} The sudden burst of management frames directing clients to disconnect is a hallmark signature for WIDS sensors.
    \item \textbf{Beacon Anomalies:} A WIDS tracks the ``fingerprint'' of legitimate APs. The sudden appearance of a second AP broadcasting the same SSID but with a different BSSID (MAC address) or different signal strength characteristics (RSSI) would trigger an immediate ``Evil Twin'' alert.
\end{itemize}

This realization underscores a critical vulnerability in many organizations that rely solely on NIDS/IPS solutions. Without a sensory layer dedicated to the RF spectrum, the initial stages of wireless exploitation remain invisible to the security operations center, regardless of how sophisticated the upper-layer packet inspection may be.

\section{Conclusion}
\label{sec:conclusion}
This project successfully demonstrated the viability of a virtualized Rogue Access Point to bypass standard user vigilance and harvest credentials. By overcoming significant hardware compatibility challenges through virtualization, the team utilized \textit{Wifipumpkin3} to deploy a functional Evil Twin.

The most significant takeaway, however, lies in the failure of the standard NIDS configuration to detect the intrusion. This underscores a critical security gap: traditional network security monitoring tools are often insufficient against wireless-specific threats. Rogue AP attacks exploit the physical layer of the network and the trust of the user, operating ``below'' the radar of many packet-inspection engines.

Consequently, this research highlights the urgent need for organizations to deploy dedicated wireless monitoring solutions and to enforce strict mutual authentication protocols (such as WPA2/3-Enterprise with certificate validation) to prevent devices from connecting to unauthorized APs. Furthermore, user education remains paramount; students and employees must be trained to verify network integrity before entering credentials, as the ``lock'' icon on a browser does not guarantee the legitimacy of the underlying network infrastructure.

\end{document}